\documentclass[11pt]{article}
\usepackage[margin=1.1in]{geometry}
\usepackage{amsmath,amssymb,amsthm,booktabs,graphicx,xcolor,hyperref,url,microtype}
\usepackage[numbers,sort&compress]{natbib}
\hypersetup{colorlinks=true,linkcolor=blue!50!black,citecolor=blue!50!black,urlcolor=blue!50!black}
\makeatletter
\renewcommand\paragraph{\@startsection{paragraph}{4}{\z@}%
  {3.25ex \@plus1ex \@minus.2ex}{-1em}%
  {\normalfont\normalsize\itshape}}
\makeatother
\newtheorem{theorem}{Theorem}
\newtheorem{proposition}[theorem]{Proposition}

\theoremstyle{remark}
\newtheorem{remark}[theorem]{Remark}

\newcommand{\E}{\mathbb{E}}
\newcommand{\dd}{\,\mathrm{d}}
\newcommand{\eps}{\varepsilon}
\newcommand{\diag}{\operatorname{diag}}
\newcommand{\rl}{\textsf{regimelib}}
\newcommand{\ql}{\textsf{QuantLib}}

\title{Security and Derivative Pricing with Fast Regime Switches}
\author{Peter Cotton\thanks{\texttt{peter.cotton@microprediction.com}. Code: \url{https://github.com/microprediction/regimelib}; documentation: \url{https://regimeswitching.org}.}}
\date{First version: September 25, 2026. This version: October 30, 2026.}

\begin{document}
\maketitle

\begin{abstract}
A regime-switching model lets any parameter of a standard model take a different value in each state of a
hidden Markov chain. For the affine and exponential-L\'evy models in common use the pricing problem reduces to
one linear system, $a' = (Q + \diag g(t))\,a$, with the chain's generator $Q$ and a model-specific forcing $g$ as
its only ingredients. We expand its solution in the mean holding time of the chain to any order, with an initial
layer for the starting regime, and identify the first correction as a Green--Kubo matrix from the group inverse
of $Q$. The same system prices bonds, swaptions, caps, credit default swaps, Asian and digital options, and equity
options under stochastic rates, with the Gil--Pelaez integrals conditioned on the regime at expiry where a payoff
depends on it; barrier, American and Bermudan instruments go to the coupled equations on a grid. Every numerical claim is checked against \ql{} in the limit where the regimes coincide and against an
independent referee with switching on, and the implementation is public.
\end{abstract}

\section{Introduction}

Hamilton's regime-switching model \citep{hamilton1989} put a hidden finite-state Markov chain behind the
parameters of a time series, and \citet{naik1993} priced options with the volatility driven by such a chain. The
pricing problem under a regime-switching model is a coupled system, one equation per regime, and for the affine
and exponential-L\'evy models in use the state dependence factors out, leaving a linear system of ordinary
differential equations in which the chain and the model meet only through the generator and a forcing term. The
reduction, and the asymptotic expansion of its solution when the chain is fast in the mean holding time, were
carried out for the Ornstein--Uhlenbeck bond and the survival probability in the author's thesis
\citep{cotton2001}. The two-state reduction for the Vasicek bond appears again in \citet{elliott2002}, and
American options with switching in \citet{buffington2002}.

\citet{yin2009option} expanded the coupled
Black--Scholes equations in the holding time of a fast chain, with the averaged volatility at leading order and
uniform error bounds, and \citet{basu2009} corrected the risk-minimising option price for a fast and for a slow
chain. The general theory of two-time-scale Markov chains, forward and backward equations without a potential
term, is \citet{yin2013} and \citet{yin2009backward}; \citet{nguyen2011,nguyen2013} carried the same expansion,
with a terminal layer and uniform error bounds, to the backward equation of a switching diffusion, where the
regime enters through the diffusion operator rather than a potential. The parallel expansion for fast stochastic
volatility, with a chain among the admissible drivers, is \citet{sircar1999}, and for a fast diffusion driving the
volatility \citet{khasminskii2005}.

The reduced system has one form for every model in \ql{} \citep{quantlib}: a forcing $g_i(t)$ per regime. Its
solution expands in the mean holding time to any order, with an initial layer, and the terms are the averaged
model at order zero, a Green--Kubo matrix of the forcing's fluctuations at order one, and a memory of the starting
regime through the group inverse of the generator. The series is asymptotic. Where it fails, the reduced system
is solved numerically instead, and the engine reports which.

The securities and derivatives a library must price follow from the same system: bonds and claims on the regime,
European and digital options, options on bonds and swaps, caps and floors, geometric Asian options, credit default
swaps and first-to-default baskets, equity options under stochastic rates, and on a grid the barrier, American and
Bermudan instruments. Two constructions in that list appear to be new. Gil--Pelaez integrals conditioned on
the regime at expiry make Jamshidian's decomposition \citep{jamshidian1989} work regime by regime, and a discounted
characteristic function prices equity options when the rate and the underlying share the chain.

The implementation, \rl{}, follows \ql{}'s classes with a chain in front of the parameters. Every model reduces
to \ql{}'s when the regimes coincide, and a suite of certificates checks that limit against \ql{} and the
switching case against a numerical, grid or Monte Carlo referee. The tables are written by a script in the
repository.

\section{The reduced system}
\label{sec:reduced}

Let $y_t$ be a continuous-time Markov chain on $\{1,\dots,n\}$ with generator $Q$ (rows sum to zero, off-diagonal
entries nonnegative), stationary distribution $\pi$, and let $x_t$ be the state of a model whose parameters depend
on $y_t$. For a price of the form $u_i(t,x) = \E[\,\Phi(x_T)\,e^{-\int_t^T r_s \dd s} \mid x_t = x, y_t = i\,]$ the
pricing equations are coupled through the generator,
\begin{equation}
  \partial_t u_i + L_i u_i + \sum_j Q_{ij}\, u_j = 0, \qquad u_i(T,\cdot) = \Phi,
  \label{eq:coupled}
\end{equation}
with $L_i$ the generator of $x$ in regime $i$ less the discounting. When the model is affine in $x$ and the payoff
exponential-affine, $u_i(t,x) = \exp(-B(\tau)\cdot x)\,a_i(\tau)$ with $\tau = T - t$, $B$ solving the model's
Riccati equation (regime-free when only drift levels, volatilities, correlations or jump intensities switch), and
$a$ solving
\begin{equation}
  a'(\tau) = \big(Q + \diag g(\tau)\big)\,a(\tau), \qquad a(0) = a_0,
  \label{eq:reduced}
\end{equation}
where $a_0 = \mathbf{1}$ for a payoff paid in every regime and $a_0 = e_j$ for one paid only if $y_T = j$.
Everything model-specific is in the forcing $g_i(\tau)$, and Table~\ref{tab:forcing} lists it for the models in
the library. Characteristic functions are the same statement with a complex payoff exponent: for the martingale
log return $X_T$ of an equity model, $\E[e^{iuX_T}\mid y_0 = i] = c(u)\,a_i(T)$ with the forcing evaluated at $u$.

\begin{table}[t]
\centering\footnotesize
\begin{tabular}{lll}\toprule
Model & switching parameters & forcing $g_i(\tau)$ \\\midrule
Vasicek $dr = a(b - r)\dd t + \sigma \dd W$ & $b,\ \sigma$ & $-a b_i B + \tfrac12 \sigma_i^2 B^2$, $B = (1-e^{-a\tau})/a$ \\
\quad with a terminal exponent $e^{-cx_T}$ & & same with $B_c = c e^{-a\tau} + (1 - e^{-a\tau})/a$ \\
Vasicek with exponential jumps & $b,\ \sigma,\ \lambda$ & $\ldots + \lambda_i\big((1 + mB)^{-1} - 1\big)$ \\
CIR $dr = k(\theta - r)\dd t + \sigma\sqrt r \dd W$ & $\theta$ & $-k\theta_i B$, $B$ the CIR Riccati solution \\
Hull--White $r = x + \varphi(t)$ & $\sigma$ & $\tfrac12\sigma_i^2 B^2$, $\varphi$ fitted with $\bar\sigma^2 = \pi\cdot\sigma^2$ \\
G2++ $r = x + y + \varphi(t)$ & $\sigma,\ \eta,\ \rho$ & $\tfrac12(\sigma_i^2 B_a^2 + \eta_i^2 B_b^2 + 2\rho_i\sigma_i\eta_i B_a B_b)$ \\
Black--Scholes, log return & $\sigma$ & $-\tfrac12\sigma_i^2 (u^2 + iu)$ \\
Heston, long-run variance & $\theta$ & $\kappa\theta_i D(\tau;u)$, $D$ the Heston Riccati solution \\
Merton, Bates & $\sigma$ or $\theta$, $\lambda$ & $\ldots + \lambda_i(\phi_J(u) - 1 - iu\,\bar\kappa)$ \\
Variance gamma & $\sigma,\ \nu,\ \theta$ & $-\nu_i^{-1}\log(1 - iu\theta_i\nu_i + \tfrac12\sigma_i^2\nu_i u^2) + iu\,\omega_i$ \\
Geometric Asian (Black--Scholes) & $\sigma$ & $iu(\mu - \tfrac12\sigma_i^2)\tfrac{\tau}{T} - \tfrac12 u^2\sigma_i^2 \tfrac{\tau^2}{T^2}$ \\
Intensity basket $\lambda^1 + \cdots + \lambda^K$ & each name's & sum of the names' forcings \\
Equity with rates (discounted c.f.) & both parts' & $g^{\text{rates}}_i(\tau; c = 1 - iw) + g^{\text{equity}}_i(w)$ \\
\bottomrule
\end{tabular}
\caption{The forcing of the reduced system \eqref{eq:reduced} for the models in \rl{}. $B$, $D$ are the
regime-free Riccati solutions of the frozen model; the prefactor $\exp(-B\cdot x_0)$ or $\exp(D v_0)$ is common to
all regimes. The Asian forcing is in time to maturity, where the weight $(1 - t/T)$ of the log price becomes
$\tau/T$.}
\label{tab:forcing}
\end{table}

\begin{proposition}[Exact reductions]
For every model in Table~\ref{tab:forcing} the coupled system \eqref{eq:coupled} is solved by
$\exp(-B(\tau)\cdot x)\,a(\tau)$ with $a$ from \eqref{eq:reduced}. The reduction fails when a switched parameter
multiplies an operator that does not commute with the rest of $L_i$: the CEV diffusion $\sigma_i^2 S^{2\beta}\partial_{SS}$
against the drift, or the Heston volatility of variance $\xi_i$ against the mean reversion.
\end{proposition}

The two-state case of the reduction, a pair of linear ODEs, is in \citet{cotton2001} and again in \citet{elliott2002} for the Vasicek bond.
One form covers the whole catalogue, including the terminal exponents that bond options need and the complex
forcings of characteristic functions. The models outside the catalogue are exactly those with a non-commuting
switched operator, and they are priced at first order.

\section{The fast-switching expansion}
\label{sec:expansion}

Write $Q = Q_1/\eps$ with $\eps = n/(-\operatorname{tr} Q)$ the mean holding time of the chain and $Q_1$ of unit
scale, and expand $a = \sum_k \eps^k a^{(k)}$. Substituting into \eqref{eq:reduced} and matching powers gives at
each order an equation $Q_1 w = f$, which has a solution only when $\pi\cdot f = 0$ (the Fredholm alternative
for the singular matrix $Q_1$); the solvability conditions determine the projections onto $\mathbf 1$, the
particular solutions are given by the group inverse $Q_1^{\#}$ \citep{meyer1975}, and the initial condition
$a(0) = a_0$, which the outer expansion cannot satisfy regime by regime, is met by an initial layer in the fast
time $\tau/\eps$. The construction is standard singular perturbation \citep{yin2013,nguyen2013}; the potential term $\diag g$ is what
separates the reduced system from the backward equations treated there. The terms have a form that can be stated.

\begin{theorem}[Expansion with initial layer]
\label{thm:expansion}
Let $g$ be smooth on $[0,T]$ and $Q$ irreducible. For every $N$ there are functions $a^{(0)},\dots,a^{(N)}$ and
layer terms $\ell^{(0)},\dots,\ell^{(N)}$, each computable from $Q_1^{\#}$, $\pi$ and $g$ by quadrature, such that
\[
  a(\tau) = \sum_{k=0}^{N} \eps^k\big(a^{(k)}(\tau) + \ell^{(k)}(\tau/\eps)\big) + O(\eps^{N+1})
\]
uniformly on $[0,T]$, with $\ell^{(k)}$ decaying exponentially in $\tau/\eps$. The leading terms are
\begin{align}
  a^{(0)}_i(\tau) &= \exp\Big(\int_0^\tau \bar g\Big), \qquad \bar g = \pi\cdot g, \label{eq:order0}\\
  a^{(1)}_i(\tau) &= a^{(0)}(\tau)\Big(\int_0^\tau K(s)\dd s \;-\; m_i(\tau)\Big), \label{eq:order1}\\
  K(s) &= -\,\pi\cdot\big(\tilde g(s)\odot Q_1^{\#}\tilde g(s)\big), \qquad m_i(\tau) = \big(Q_1^{\#}\tilde g(\tau)\big)_i, \notag
\end{align}
where $\tilde g = g - \bar g\mathbf 1$ is the centred forcing and $\odot$ the entrywise product; $\ell^{(0)} = 0$ and
the first layer is $\ell^{(1)}(s) = e^{Q_1 s}\,Q_1^{\#}\tilde g(0)$, which cancels $-m(0)$ at $\tau = 0$ and decays
because $Q_1^{\#}\tilde g(0)$ has no component along $\mathbf 1$ ($\pi Q_1^{\#} = 0$).
\end{theorem}

The order-zero term is the averaged model: the forcing replaced by its stationary average, which for Vasicek is
the bond of the averaged mean level and averaged variance. The first correction has two parts. The integral of
$K$ is a Green--Kubo term: for a vector forcing $g = \sum_j \phi_j f_j$ with regime-dependent coefficients $f_j$
it is the quadratic form $\sum_{jk}\phi_j\phi_k K_{jk}$ in the matrix
\begin{equation}
  K_{jk} = \int_0^\infty \operatorname{Cov}\big(f_j(y_0), f_k(y_t)\big)\dd t = -\,\pi\cdot\big(\tilde f_j \odot Q_1^{\#}\tilde f_k\big),
  \label{eq:gk}
\end{equation}
the integrated autocovariance of the parameters along the chain, the extra convexity the switching creates. Only
its symmetric part enters a scalar forcing. The antisymmetric part, nonzero for a non-reversible chain, enters
when the switched operators do not commute.

The memory term $m_i$ is the group inverse applied to the centred forcing, evaluated in the starting regime. It
distinguishes the price seen from regime $i$ from the price seen from regime $j$, and it decays with the layer as
the chain forgets its start. Higher orders follow by the same recursion, and the library computes them to any
order, with the integrals done exactly when $g$ is an exponential sum and by Chebyshev series otherwise.

The certificate
\texttt{verify\_theorem.py} in the repository evaluates the residual of the order-one approximation, layer included,
for a three-regime chain with the Vasicek forcing and with a constant forcing: the supremum over $[0,T]$ divided by
$\eps^2$ is constant to within a few percent across $\eps = 0.2, 0.1, 0.05, 0.025$, and without the layer the
residual for the constant forcing is $0.043\,\eps$, so the layer is not optional.

\begin{remark}[The series is asymptotic]
The expansion parameter is not $\eps$ alone but $\eps\,\|g\|$, and for a characteristic function $g$ grows with the
Fourier variable, $|g_i(u)| \sim \tfrac12\sigma_i^2 u^2$ for Black--Scholes. At fixed $\eps$ the terms therefore
stop decreasing beyond some order, and at fixed order they diverge beyond some frequency. The engine
therefore reports the size of the last term kept and stops at the best truncation when asked. At Fourier nodes
where the last term is not smaller than the one before it, the reduced system is solved numerically instead of
expanded, and the price is a hybrid: the expansion where it converges, the numerical solution where it does not.
The geometric Asian option, whose characteristic function decays slowly in $u$, needs the hybrid at moderate
holding times; replacing the diverging nodes by the averaged value leaves an error of $10^{-3}$ there.
\end{remark}

\begin{proposition}[Two regimes, constant forcing]
For $n = 2$ and constant $g$ the solution of \eqref{eq:reduced} is exact: with $M = Q + \diag g$ and
eigenvalues $\mu_\pm = \tfrac12(\operatorname{tr}M \pm \sqrt{\operatorname{tr}^2 M - 4\det M})$,
$a(\tau) = \big[e^{\mu_+\tau}(M - \mu_- I) - e^{\mu_-\tau}(M - \mu_+ I)\big]\mathbf 1/(\mu_+ - \mu_-)$.
This is the closed-form characteristic function of every two-regime Black--Scholes, Merton and variance-gamma
model, and the expansion of Theorem~\ref{thm:expansion} is its expansion in $\eps$.
\end{proposition}

\section{Securities and derivatives}
\label{sec:instruments}

Table~\ref{tab:instruments} lists the securities and derivatives the library prices and the route from the
reduced system to each: what is solved, with which terminal vector, and what is then integrated or projected.
Payoffs of the terminal state go through the reduced system and one integral. Payoffs of the path go to the
coupled equations on a grid.

\begin{table}[t]
\centering\footnotesize
\resizebox{\textwidth}{!}{\begin{tabular}{lll}\toprule
Instrument & models & route \\\midrule
Zero-coupon bond; face paid only in regime $j$ & rates, intensities & reduced system, $a_0 = \mathbf 1$ or $e_j$ \\
Coupon bond & rates & sum of zero-coupon bonds \\
European call, put & equity, equity with rates & characteristic function, Lewis integral \\
Cash- and asset-or-nothing digital & equity & characteristic function, Gil--Pelaez integral \\
Geometric Asian option & Black--Scholes & reduced system with the $\tau/T$ forcing, Lewis integral \\
Option on a zero-coupon bond & Vasicek, Hull--White, G2++ & terminal exponent $c - iu$, $a_0 = e_j$, Gil--Pelaez per regime \\
Option on a coupon bond, European swaption & same & Jamshidian's decomposition regime by regime \\
Cap, floor & same & puts, calls on the period bonds \\
Credit default swap, first-to-default & intensities & survival probabilities at the schedule dates \\
Barrier option & equity & coupled equations on a grid, one block per regime \\
American option, Bermudan swaption & equity; rates & the same, projected onto the exercise value \\
\bottomrule
\end{tabular}}
\caption{What \rl{} prices and how. Every model of Table~\ref{tab:forcing} supplies the forcing; the terminal
vector carries the regime at expiry; one integral or one grid completes the price.}
\label{tab:instruments}
\end{table}

\paragraph{Bonds and claims on the regime.} A zero-coupon bond is $\exp(-B(T)\cdot x_0)\,a_i(T)$ with
$a_0 = \mathbf 1$. With $a_0 = e_j$ the same solve prices the face value paid only if $y_T = j$, the memory of
the regime as a security, and the sum over $j$ recovers the bond. A coupon bond is the sum of its zero-coupon
bonds, and when the state is a default intensity the bond is the survival probability.

\paragraph{European and digital options on equity.} European options use Lewis's formula \citep{lewis2001} with
the reduced characteristic function,
\begin{equation}
  C = e^{-rT}\Big[F - \frac{\sqrt{FK}}{\pi}\int_0^\infty \operatorname{Re}\big(e^{iuk}\,\phi(u - \tfrac i2)\big)\frac{\dd u}{u^2 + \tfrac14}\Big],
  \qquad \phi(z) = c(z)\,a_i(T;z),
  \label{eq:lewis}
\end{equation}
$k = \log(F/K)$, with the frequency cutoff found from the averaged characteristic function, which every model
exposes in closed form. A cash-or-nothing call paying one if $S_T > K$ is the Gil--Pelaez integral of the same
function,
\begin{equation}
  e^{-rT}\Big[\tfrac12 + \frac1\pi\int_0^\infty \operatorname{Re}\Big(\frac{e^{iuk}\,\phi(u)}{iu}\Big)\dd u\Big],
  \label{eq:digital}
\end{equation}
and an asset-or-nothing option is the same integral under the share measure, $\phi(u - i)/\phi(-i)$ in place of
$\phi(u)$ and the forward in front. Differentiating under the integral gives delta, gamma, theta and rho, and for
stochastic-volatility models the vega in the initial variance; theta uses $\partial_T\phi$, which the reduced
system supplies as $(Q + \diag g(T))\,a(T)$ without a further solve.

\paragraph{Options on bonds, swaptions, caps and floors.}
Under a Gaussian short rate the bond at expiry $T$ for maturity $S$ in regime $j$ is $A_j e^{-b x_T}$ with
$A_j = a_j(S - T)$ from the reduced system and $b = B(S - T)$, so the option pays $(A_j e^{-bx_T} - K)^+$ on the
event $\{x_T < x^*_j\}$ with one crossing $x^*_j = \log(A_j/K)/b$ per regime. Each term is a discounted
probability of the form $\E[e^{-\int_0^T r}\,e^{-c x_T}\,\mathbf 1\{x_T < x^*_j\}\,\mathbf 1\{y_T = j\}]$. Write
$\Psi_j(c - iu)$ for the reduced system with the terminal exponent $c - iu$ and $a_0 = e_j$; Gil--Pelaez inversion
gives
\begin{equation}
  \E\big[e^{-\int_0^T r} e^{-cx_T}\mathbf 1\{x_T < x^*_j\}\mathbf 1\{y_T = j\}\big]
  = \tfrac12\Psi_j(c) - \frac1\pi\int_0^\infty \operatorname{Im}\big(e^{-iux^*_j}\,\Psi_j(c - iu)\big)\frac{\dd u}{u}.
  \label{eq:gilpelaez}
\end{equation}
The regime indicator at expiry is carried by the initial vector $e_j$ and costs nothing beyond one solve per
regime; the integrals are done on Kronrod panels with an error estimate and halved until it is met.

A coupon bond $\sum_k c_k A_{kj}e^{-b_k x}$ is monotone in $x$ in every regime, so Jamshidian's decomposition
\citep{jamshidian1989} holds regime by regime: one crossing $x^*_j$ per regime, and the option is a sum of
zero-coupon bond options with strikes $A_{kj}e^{-b_k x^*_j}$. The strikes differ by regime, which is the only
change from the model without regimes. A receiver swaption is the call on the coupon bond at par and a payer
swaption the put. A caplet on $[t_{k-1}, t_k]$ with strike $K$ is $1 + \tau_k K$ puts on the bond maturing at
$t_k$, struck at $1/(1 + \tau_k K)$ and expiring at $t_{k-1}$, and a floorlet is the same number of calls. Under
Hull--White the deterministic part of the discounting scales the cash flows; under G2++ the same construction
applies to the zero-coupon bond with $z = B_a x_T + B_b y_T$ as the single variable and a two-factor terminal
forcing.

\paragraph{Observed and inferred regimes.}
A price in regime $i$ is the price when the regime is known at the valuation date. From a belief $p$ about it, a
claim whose payoff depends only on observed quantities (a bond, an option on an equity, a default) is worth
$\sum_i p_i u_i$, because nobody acts on the regime and pricing is linear. An option on a bond or a swap is
different: its payoff involves the bond price at expiry, and that price depends on what is then known.

The construction with one exercise boundary per regime and the bond at expiry equal to $A_j e^{-b x_T}$ on
$\{y_T = j\}$ is the price when the regime is observed. It is also the price when the regime is inferred from the
path, provided a diffusion coefficient differs between every two regimes, since the quadratic variation of $x$ then
reveals $y$ at once. When only a drift level switches the regime is not revealed; the bond at expiry is the
belief-weighted $\sum_j p^j_T A_j e^{-b x_T}$, and by Jensen's inequality the option is worth less than under
observation.

For two regimes the belief $p_t = \mathbb{P}(y_t = 1 \mid x_s,\ s \le t)$ follows the filter of
\citet{wonham1965},
\[
  \dd p_t = \big(q_{21}(1 - p_t) - q_{12}\,p_t\big)\dd t + p_t(1 - p_t)\,\frac{a\,(b_1 - b_2)}{\sigma}\dd\nu_t,
\]
with $\nu$ the innovation, which also drives $\dd x_t = a(p_t b_1 + (1 - p_t)b_2 - x_t)\dd t + \sigma\dd\nu_t$. The
pair $(x, p)$ is Markov, the bond at $(x, p)$ is known in closed form, and the option solves one equation in
$(x, p)$ with no regime blocks. The same distinction applies to early exercise, where the observed-regime price is
an upper bound. The library computes both, and with level-only switching the two differ by about fifteen percent
for a one-into-five payer swaption on a chain with a two-year holding time.

\paragraph{Credit default swaps.}
When the state is a default intensity, $Q_i(t) = e^{-B(t)x_0}\,a_i(t)$ is the survival probability from regime
$i$. A swap paying the spread $s$ at dates $t_1 < \dots < t_n$, with the accrued premium paid at default and
recovery $R$, has the two legs
\begin{align}
  \text{premium} &= s\sum_k\Big[\tau_k D(t_k)\,Q(t_k) + \tfrac12\tau_k D(t_k^m)\big(Q(t_{k-1}) - Q(t_k)\big)\Big], \notag\\
  \text{protection} &= (1 - R)\sum_k D(t_k^m)\big(Q(t_{k-1}) - Q(t_k)\big),
  \label{eq:cds}
\end{align}
with $\tau_k = t_k - t_{k-1}$, $t_k^m$ the midpoint of the period and $D$ the risk-free discount, which is the
mid-point convention of \ql{}'s engine. The fair spread is the ratio of the protection leg to the premium leg
per unit spread. Only survival probabilities enter, so the closed forms of Section~\ref{sec:expansion} apply date
by date: for a two-regime Vasicek intensity with mean levels 6\% and 1\% on a chain with rate 5, the second-order
closed form gives a fair spread of 178.14 basis points and the engine 178.14, agreeing to six parts in a million.

Several names on one chain have joint survival equal to the bond of the summed forcing, so a first-to-default
swap prices as a single-name swap on the basket, and the default correlation the common regime induces is read
off the joint and marginal survivals.

\paragraph{Geometric Asian options.}
Under a Black--Scholes equity with a switching volatility the continuous geometric average is $G = S_0 e^Y$ with
$Y = \int_0^T (1 - t/T)\dd\log S_t$: each increment of the log price enters scaled by the fraction of the life
that remains. In time to maturity that fraction is $\tau/T$, so $\E[e^{iuY}\mid y_0 = i] = a_i(T)$ with the
polynomial forcing of Table~\ref{tab:forcing}, every integral of the expansion is elementary, and the order-zero
term is the Kemna--Vorst result that the average of a lognormal path has half the drift and a third of the
variance. The option is Lewis's formula \eqref{eq:lewis} on $G$, whose forward is $S_0\,\E[e^Y]$. On a two-regime
volatility of 30\% and 15\% switching at rate 50, the second-order closed form and the engine agree to under
$10^{-6}$ relative at strikes 95 and 105.

\paragraph{Equity with stochastic rates.}
For an equity and a short rate driven by one chain with independent Brownian motions, the discounted
characteristic function $\Psi(w) = \E[e^{-\int_0^T r}\,S_T^{iw}]$ factors, given the regime path, into the rates
part $\E[e^{(iw - 1)\int r}]$, the Gaussian terminal formula with weight $c = 1 - iw$, and the equity's
martingale characteristic function, so the reduced system takes the sum of the two forcings
(Table~\ref{tab:forcing}). Lewis's formula in discounted form,
$C = S_0e^{-qT} - \frac{\sqrt K}{\pi}\int_0^\infty \operatorname{Re}\big(K^{-iu}\Psi(u - \tfrac i2)\big)\frac{\dd u}{u^2 + 1/4}$,
prices the option with the discount and the return dependent through the regime. An equity--rate correlation is
admitted for the Black--Scholes equity, where the exponent stays Gaussian given the path and the cross term
$-c\,iw\,\rho\,\sigma_S\sigma_r B(\tau)$ joins the forcing. In the frozen limit these agree with \ql{}'s analytic
Black--Scholes--Hull--White and Heston--Hull--White engines to $10^{-8}$ and $10^{-6}$ relative.

\paragraph{Barrier, American and Bermudan instruments.}
These depend on the path of the state and not only on its terminal value, so the state does not factor out and
there is no reduced system. They are priced on the coupled system \eqref{eq:coupled} by Crank--Nicolson after
Rannacher's implicit start, on a grid in the state with one block per regime. A knock-out barrier is the
condition $u_i = \text{rebate}$ at the barrier in every regime, and a knock-in is the vanilla less the
knock-out. An American option is the variational inequality
\begin{equation}
  \max\Big(\partial_t u_i + L_i u_i + \sum_j Q_{ij}u_j,\ \Phi - u_i\Big) = 0,
  \label{eq:american}
\end{equation}
with an exercise boundary that differs by regime.

A Bermudan swaption projects onto the exercise value at each date, and that value is the coupon bond in each
regime from the reduced system, evaluated on the grid. Two-factor models (G2++, a switched volatility of
variance, equity with rates) use a tensor grid, uniform or concentrated by the sinh map of \citet{tavella2000}.
On a two-regime Black--Scholes put with volatilities 30\% and 15\%, the down-and-out and down-and-in prices,
1.1979 and 5.9760, add to the European price 7.1739 to four decimals, and the American put is 7.7214.

\section{Models outside the exact reduction}
\label{sec:firstorder}

When a switched parameter multiplies an operator $A_j$ that does not commute with the averaged generator
$\bar L$, the pricing equation is $\partial_t u_i + (\bar L + \sum_j \tilde f_j(i) A_j)u_i + \sum_k Q_{ik}u_k = 0$ and
the expansion above still applies to the operator-valued system, with the first
correction now
\begin{equation}
  u \approx e^{T\bar L}u_0 + \int_0^T e^{(T-s)\bar L}\Big(\sum_{jk} K_{jk}A_jA_k\Big)e^{s\bar L}u_0\dd s - \sum_j m_j(i)\,A_j\,e^{T\bar L}u_0,
  \label{eq:firstorder}
\end{equation}
the Green--Kubo matrix \eqref{eq:gk} applied through Duhamel's formula, and the memory term acting on the averaged
solution. The antisymmetric part of $K$ now contributes through the commutators $[A_j, A_k]$, which is how a
non-reversible chain leaves a trace that no reversible chain can. The library evaluates \eqref{eq:firstorder} on
a grid by an augmented linear system exponentiated once, reports the relative size of the correction, and
estimates the neglected second-order term as its square; on the CEV model with a switched volatility the
estimate tracks the actual error against the coupled system to within a factor of two across chain speeds, and
on Heston with a switched volatility of variance the first-order price is within $10^{-4}$--$7\times10^{-4}$ of the
coupled two-dimensional PDE while the correction itself is 0.3--3\% of the price.

\section{Parameter sensitivities}
\label{sec:greeks}

Because the chain enters the first-order price only through $\pi$, $K$ and $m$, sensitivities to the model's
parameters are the chain rule through those numbers, and their derivatives are exact.

\begin{proposition}
Let $dQ$ be a perturbation of the generator with zero row sums. Then
\[
  d\pi = -\pi\, dQ\, Q^{\#}, \qquad
  dQ^{\#} = -Q^{\#}dQ\,Q^{\#} + \mathbf 1\pi\, dQ\,(Q^{\#})^2 + (Q^{\#})^2 dQ\,\mathbf 1\pi,
\]
and $K$ and $m$ vary accordingly: $dK_{jk} = -\big[d\pi\cdot(\tilde f_j\odot Q^{\#}\tilde f_k) + \pi\cdot(d\tilde f_j\odot Q^{\#}\tilde f_k) + \pi\cdot(\tilde f_j\odot dQ^{\#}\tilde f_k + \tilde f_j\odot Q^{\#}d\tilde f_k)\big]$ with $d\tilde f_j = -(f_j\cdot d\pi)\mathbf 1$.
For a perturbation of a per-regime coefficient $f_j(i)$, $K$ is bilinear and $m$ linear in the forcing, so the
derivatives are $K$ and $m$ evaluated on unit vectors.
\end{proposition}

With the first-order bond as a symbolic expression in $\pi\cdot\theta$, $\pi\cdot\sigma^2$, $K$ and $m$,
$\partial P/\partial\theta_i$, $\partial P/\partial\sigma_i$ and $\partial P/\partial Q_{ab}$
are formulas rather than bumps; Table~\ref{tab:greeks} checks them against finite differences of the engine at
the same order.

\section{Numerical certificates}
\label{sec:numerics}

Table~\ref{tab:frozen} is the frozen limit: two regimes with identical parameters, the chain still switching,
priced by \rl{} and by \ql{}'s engine for the same instrument. The characteristic-function engines agree with
\ql{} to quadrature accuracy; the grid engines carry their discretisation error; the credit default swap
differs by $9\times10^{-5}$, which is the rounding of \ql{}'s schedule to whole days.

Table~\ref{tab:orders} is the expansion with switching on: a payer swaption under a two-regime Vasicek model
(mean levels 6\% and 2\%, volatilities 1.5\% and 0.8\%) at each order against the numerical solution of the
reduced system, for four holding times. The error falls with the order at every holding time, though not
monotonically: order three is worse than order two throughout. For the faster chains the order-six error has
reached the tolerance of the numerical reference. The adaptive engine stops at the order where the last term
stops shrinking.

Table~\ref{tab:timings} gives timings on a laptop. For a bond, one evaluation of the series, the expansion is
the fastest route. For a Fourier integral the numerical engine solves the reduced system for every node in a
single batched call, and the series evaluated node by node at order four costs more; the expansion supplies the
formula, the diagnostics and the greeks under the integral, not speed. The Monte Carlo referee is pure Python
and priced per path, and is a referee rather than an engine.

Table~\ref{tab:greeks} gives the parameter sensitivities.

\begin{table}[t]\centering\footnotesize
\resizebox{\textwidth}{!}{\begin{tabular}{llrrr}\toprule
Instrument & QuantLib engine & regimelib & QuantLib & rel.\ diff.\\\midrule
Vasicek bond, $T=5$ & \texttt{Vasicek::discountBond} & 0.83445782 & 0.83445782 & 0.0e+00\\
Hull--White bond call & \texttt{HullWhite::discountBondOption} & 0.014645681 & 0.014645681 & 2.4e-15\\
G2++ bond call & \texttt{G2::discountBondOption} & 0.016784068 & 0.016784068 & 2.8e-13\\
Hull--White payer swaption $2\times5$ & \texttt{JamshidianSwaptionEngine} & 0.0084708386 & 0.0084708405 & 2.2e-07\\
Hull--White Bermudan swaption & \texttt{FdHullWhiteSwaptionEngine} & 0.013171548 & 0.013171543 & 3.9e-07\\
Hull--White cap, 4 caplets & \texttt{AnalyticCapFloorEngine} & 0.013715712 & 0.013715712 & 3.3e-15\\
Black--Scholes put & \texttt{AnalyticEuropeanEngine} & 11.504233 & 11.504233 & 8.7e-13\\
Heston put & \texttt{AnalyticHestonEngine} & 9.2806468 & 9.2806468 & 1.1e-11\\
American put & \texttt{FdBlackScholesVanillaEngine} & 11.793526 & 11.793912 & 3.3e-05\\
Down-and-out call & \texttt{AnalyticBarrierEngine} & 9.5800708 & 9.5800332 & 3.9e-06\\
Geometric Asian call & \texttt{AnalyticContinuousGeometric...} & 5.8327546 & 5.8327546 & 1.7e-12\\
Black--Scholes--Hull--White put, $\rho=0.4$ & \texttt{AnalyticBSMHullWhiteEngine} & 14.522982 & 14.522982 & 1.3e-12\\
CDS fair spread, CIR intensity & \texttt{MidPointCdsEngine} & 0.015712784 & 0.015714177 & 8.9e-05\\
\bottomrule\end{tabular}
}
\caption{Frozen limit: two identical regimes on a chain with rates 3 and 5 per year, against \ql{}. Grid engines
(American, Bermudan, barrier) carry the grid's discretisation error; the characteristic-function engines agree to
quadrature accuracy.}
\label{tab:frozen}
\end{table}

\begin{table}[t]\centering\small
\begin{tabular}{lrrrrrr}\toprule
$\varepsilon$ & order 0 & 1 & 2 & 3 & 4 & 6\\\midrule
0.050 & 7.2e-03 & 4.5e-05 & 1.1e-07 & 1.4e-07 & 1.9e-09 & 8.5e-11\\
0.100 & 1.5e-02 & 1.7e-04 & 9.9e-07 & 2.5e-06 & 1.9e-08 & 1.9e-08\\
0.200 & 3.1e-02 & 5.5e-04 & 1.5e-05 & 4.6e-05 & 3.1e-06 & 2.6e-10\\
0.333 & 5.3e-02 & 5.9e-03 & 1.6e-04 & 3.8e-04 & 1.3e-06 & 9.4e-09\\
\bottomrule\end{tabular}

\caption{Relative error of the expansion at each order against the numerical solution of the reduced system, for
a $2\times5$ payer swaption under a two-regime Vasicek model, by mean holding time $\eps$ of a symmetric two-state
chain.}
\label{tab:orders}
\end{table}

\begin{table}[t]\centering\small
\begin{tabular}{lrrr}\toprule
Instrument (switching on) & expansion, order 4 & numerical & grid / Monte Carlo\\\midrule
Vasicek bond & 7.7 ms & 8.3 ms & 46150 ms (MC, $10^5$ paths)\\
Vasicek swaption & 4687 ms & 418 ms & 55 ms (grid)\\
Black--Scholes call / American put & 233 ms & 10 ms & 56 ms (grid, American)\\
Heston call & 10273 ms & 9930 ms & --\\
\bottomrule\end{tabular}

\caption{Wall-clock time per price with switching on (two regimes, rates 12 and 8), Python on a laptop.}
\label{tab:timings}
\end{table}

\begin{table}[t]\centering\small
\begin{tabular}{lrrr}\toprule
Parameter & formula & finite difference & rel.\ diff.\\\midrule
$\partial P/\partial\theta_1$ & -0.62010531 & -0.62010531 & 1.6e-11\\
$\partial P/\partial\sigma_2$ & 0.018654226 & 0.018654226 & 1.1e-10\\
$\partial P/\partial Q_{12}$ & 0.0028187193 & 0.0028187193 & 4.2e-09\\
$\partial P/\partial Q_{32}$ & -0.0027397038 & -0.0027397038 & 5.2e-09\\
\bottomrule\end{tabular}

\caption{Parameter sensitivities of the first-order Vasicek bond under a three-regime chain, from the derivative of the group inverse
and from central differences of the engine at order one.}
\label{tab:greeks}
\end{table}

\section{Reproducibility}
\label{sec:software}

The implementation is released under the MIT licence at \url{https://github.com/microprediction/regimelib}, on
PyPI as \texttt{regimelib}, with documentation at \url{https://regimeswitching.org/python/}. A model in the
all-orders tier is a statement of its forcing; a model whose reduction is not exact goes to the first-order tier
with its diagnostics. Instruments accept \ql{}'s payoff, exercise and date objects, and the greeks are computed by
differentiating the pricing formula.

Every certificate checks the frozen limit against \ql{} and the switching case against a referee that does not
share the method: the numerical solution of \eqref{eq:reduced} for the expansion, the coupled PDE on a grid for
the first-order tier, Monte Carlo with exact regime paths for the grids.

The frozen limit alone is not enough, and two blind spots fix the design. With all regimes equal, a forcing integrates to the same value in calendar
time and in time to maturity, so the frozen limit cannot detect the direction of time in a time-dependent
forcing; with switching on, the Asian option moves by 7\% between the two. And a reversible chain has a symmetric
Green--Kubo matrix, so only a non-reversible test chain separates the symmetric part of $K$ from the whole in a
cross term, a $10^{-4}$ effect in the first-order coupon-bond formula. The certificates therefore run with
switching on and on non-reversible chains.

Closed forms are in \texttt{regimelib.symbolic}: the two-state Vasicek bond to second order, the first-order
bond under any chain for Vasicek, CIR (integrals closed by the Riccati identity $\int B^2 = \frac2{\sigma^2}(T - kI_1 - B(T))$)
and Vasicek with jumps, the exact two-state constant-forcing solution, and the parameter greeks above, each
checked against the engine at the same order to round-off. The tables are written by
\texttt{papers/regimelib/verify\_tables.py} and the theorem's certificate is \texttt{verify\_theorem.py} beside it.

\section{Conclusion}

Theorem~\ref{thm:expansion} is stated for a smooth forcing and an irreducible chain, and its proof is the
singular-perturbation argument of \citet{yin2013}; the certificate checks the order-one case
numerically and not the general one. The first-order tier has no second-order term and no certificate beyond the
error estimate against the coupled PDE. We leave to future work the second order of that tier, the equity--rate
correlation for a Heston equity, and ports of the reference implementation to other languages under the parity
contract of \texttt{regimeswitching.org}.

\bibliographystyle{plainnat}
\bibliography{regimelib}
\end{document}